\documentclass{pasj00}

\begin{document}
\SetRunningHead{Y. Takeda and M. Takada-Hidai}{[S/Fe] Behavior of Metal-Poor 
Stars}
\Received{2010/07/29}
\Accepted{2010/08/23}

\title{Exploring the [S/Fe] Behavior of Metal-Poor Stars\\
with the S~I 1.046~$\mu$m Lines
\thanks{Based on data collected at Subaru Telescope, which is operated by 
the National Astronomical Observatory of Japan.}
\thanks{The large data tables are separately provided in the machine-readable 
form as electronic tables E1 and E2.}
}

%

\author{Yoichi \textsc{Takeda}}
\affil{National Astronomical Observatory of Japan 
2-21-1 Osawa, Mitaka, Tokyo 181-8588}
\email{takeda.yoichi@nao.ac.jp}
\and 
\author{Masahide \textsc{Takada-Hidai}}
\affil{Liberal Arts Education Center, Tokai University, 
1117 Kitakaname, Hiratsuka, Kanagawa 259-1292}
\email{hidai@apus.rh.u-tokai.ac.jp}

%


\maketitle

\begin{abstract}
In an attempt of clarifying the [S/Fe] behavior with the run of 
[Fe/H] in the metal-poor regime which has been a matter of debate, 
an extensive non-LTE analysis of near-IR S~{\sc i} triplet lines 
(multiplet 3) at 1.046~$\mu$m was carried out for selected 
33 halo/disk stars in a wide metallicity range of [Fe/H] 
$\sim -3.7$ to $\sim +0.3$, based on the spectral data 
collected with IRCS+AO188 of the Subaru Telescope.
We found an evidence of considerably large [S/Fe] ratio
amounting to $\sim$~+0.7--0.8~dex at very low metallicity of 
[Fe/H]~$\sim -3$, which makes marked contrast with other
$\alpha$-elements (Mg, Si, Ca, Ti) flatly showing moderately 
supersolar [$\alpha$/Fe] of $\sim 0.3$~dex.
Meanwhile, a locally-flat tendency of [S/Fe] at $\sim +0.3$ is 
seen at $-2.5 \ltsim$~[Fe/H]~$\ltsim -1.5$. These results may 
suggest that the nature of [S/Fe] in metal-poor halo stars
is not so simple as has been argued (i.e., neither being globally 
flat independent of [Fe/H] nor monotonically increasing with a
decrease in [Fe/H]), but rather complicated with a local 
plateau around [Fe/H] $\sim -2$ followed by a discontinuous 
jump between the narrow interval of $-3 \ltsim$~[Fe/H]~$\ltsim -2.5$.
\end{abstract}

%


\section{Introduction}

Sulfur belongs to the important group of $\alpha$-capture elements, 
whose abundances in very metal-poor stars play a key role for 
studying the galactic chemical evolution, because most of them 
are considered to have been synthesized in short-lived massive 
stars and thrown out by type II supernovae at the early history 
of the Galaxy. Among these, S (along with O) deserves 
a particular attention because of its chemically ``volatile'' 
nature; i.e., it is difficult to condense into solid owing to 
its low condensation temperature ($T_{\rm c} \sim$ 650 K) 
unlike other ``refractory'' $\alpha$ group (Mg, Si, Ca, Ti) with 
high $T_{\rm c}$ of $\sim$ 1300--1500 K. 
This fact characterizes the S abundances of metal-deficient stars 
as a reliable probe for the $\alpha$-chemistry of old galactic gas 
because S atoms synthesized and emitted by SNe~II are directly 
circulated into the gas (from which stars form), in contrast to 
other refractory species which might have at first fractionated 
onto dust and later deposited to the gas after a considerable 
elapse of time (time-delayed deposition; e.g., Ramaty et al. 2001). 
For this reason, it is of paramount importance to establish 
the run of [S/Fe] with a change of [Fe/H] in very metal-poor stars.

Unfortunately, however, no consensus has yet been accomplished 
regarding the behavior of [S/Fe] of old halo stars, since two 
abundance indicators (S~{\sc i} 8693--4 lines of multiplet 6 and 
S~{\sc i} 9212/9228/9237 lines of multiplet 1), which have been 
mainly used to trace the S abundances of metal-poor stars, tend 
to yield discordant results in the very low metallicity 
regime: The former suggested an ever-increasing [S/Fe]$_{86}$ 
with a decrease of [Fe/H] up to [S/Fe]$_{86} \sim +0.8$ at 
[Fe/H] $\sim -2.5$ (``rising'' tendency; cf. Israelian \& Rebolo 2001; 
Takada-Hidai et al. 2002), while the latter resulted in a trend 
of nearly constant [S/Fe]$_{92}$ at a mildly supersolar value of 
$\sim$~+0.3--0.5 irrespective of the metallicity down to [Fe/H] 
$\sim -3$ (``flat'' tendency; cf. Ryde \& Lambert 2004; Nissen et al. 
2004; Takada-Hidai et al. 2005; Takada-Hidai \& Sargent 2005).
Since all these studies on [S/Fe]$_{92}$ invoked the assumption
of LTE, Takeda et al. (2005b) investigated the non-LTE effect 
on abundance determinations from S~{\sc i} 9212/9228/9237 lines 
and found appreciable ``negative'' corrections\footnote{
The significance of downward non-LTE corrections for these 
multiplet 1 lines has also been confirmed by recent non-LTE 
calculations done by Korotin (2009), who derived results
similar to those of Takeda et al. (2005b).} amounting to 
$\sim$~0.2--0.3~dex, which embarrassingly makes the discrepancy 
between [S/Fe]$_{86}$ and [S/Fe]$_{92}$ even larger. 
While Nissen et al. (2007b, 2008) concluded by taking into account 
the non-LTE effect that galactic halo stars distribute around 
a plateau of [S/Fe]$_{92} \sim$ +0.2--0.3 like other 
$\alpha$-elements, the controversial ``flat vs. increasing'' 
debate depending on different abundance indicators has not yet 
been settled. Do two populations showing different [S/Fe] behaviors 
(high- and low-[S/Fe]) exist in the metal-poor regime, as suggested 
by Caffau et al. (2005)? 

From a viewpoint of reliability, each of these two S abundance 
indicators actually have specific shortcomings: Although S~{\sc i} 
8693--4 lines are well-behaved in the sense that they are insensitive 
to the non-LTE effect as well as the microturbulence, their strengths 
are not so large to be reliably usable for very metal-poor stars 
at [Fe/H] $\ltsim -2$. In contrast, S~{\sc i} 9212/9228/9237 lines are 
sufficiently strong as to be invoked for studying [S/Fe] in the 
metallicity range down to [Fe/H] $\sim -3$. Unfortunately, however,
they are located in a spectral region considerably contaminated by a 
jungle of telluric H$_{2}$O lines, which (despite any effort of 
elimination) may seriously affect the accuracy of $EW$ measurements 
for very weak lines.

Considering this situation, Takeda et al. (2005b) proposed an 
alternative use of S~{\sc i} triplet lines of multiplet 3 
(4p $^{5}{\rm P}$ -- 4d $^{5}{\rm D}^{\rm o}$) at 10455--10459~$\rm\AA$
for exploring [S/Fe] of metal-poor stars, which are nearly as 
strong as S~{\sc i} 9212/9228/9237 lines and locate in a region 
almost free from any contamination of telluric lines; thus being 
certainly advantageous compared to the other two. 
Therefore, it is interesting to see which kind of [S/Fe] trend 
would result from these 10455--9 lines.
As far as we know, however, only a limited studies on [S/Fe] of
metal-poor stars has been carried out so far with these near-IR lines: 
While the first report for G~29-23 ([Fe/H] = $-1.7$) was made by 
Nissen et al. (2007a, b), Caffau et al. (2010) recently published
the results for four stars (BD$-05^{\circ}$3640, HD~140283,
HD~181743, and HD~211998) ranging from [Fe/H] $\sim -1.2$  
to $\sim -2.4$, both being based on VLT/CRIRES spectra.
Although these studies derived [S/Fe] ratios to be in the range of
+0.3--0.7, this sample is still too small, and a more extensive 
investigation is evidently needed for any statistically
meaningful information.

Hence, we decided to investigate this controversial situation on the
[S/Fe] vs. [Fe/H] trend of metal-deficient stars by ourselves
with these S~{\sc i} 10455--10459 lines.
Toward this aim, we secured moderately high-dispersion 
spectra in $zJ$-band for 33 stars of various metallicity 
in 2009 July by using IRCS+AO188 of the Subaru Telescope. 
The purpose of this paper is to report the sulfur abundances 
of these stars resulting from our analysis. 

The remainder of this article is organized as follows. After 
describing the observational data (section 2) and stellar 
parameters (section 3) of the target stars, we explain 
the details of abundance determinations in section 4, 
followed by the discussion (section 5) where the results are 
presented and examined in comparison with the published work. 
The conclusion is summarized in section 6, followed by an 
Appendix where the nature of HD~219617 (double-star system
comprising similar components) is briefly mentioned. 

\section{Observational Data}

We selected 33 targets of halo/disk stars in the 
metallicity range of $-3.7 \ltsim$~[Fe/H]~$\ltsim +0.3$, 
for which published atmospheric parameters are available.
While our halo sample ([Fe/H]~$\ltsim -1$) comprises dwarfs
as well as giants, most of the disk targets ($-1 \ltsim$~[Fe/H]) 
are dwarfs. The list of the program stars is presented in table 1.

The observations were conducted on 2009 July 29 and 30 (UT)
by using the Infrared Camera and Spectrograph (IRCS;
Kobayashi et al. 2000; Tokunaga et al. 1998) along with
the 188-element curvature-based adaptive optics system (AO188),  
which is mounted on the IR Nasmyth focus of the 8.2~m Subaru 
Telescope atop Mauna Kea. The $zJ$-band (1.04--1.19~$\mu$m) 
spectra were taken in the echelle spectrograph mode of IRCS, 
which is equipped with a Raytheon 1024$\times$1024 InSb array 
with an Aladdin II multiplexer. We used a very narrow slit of 
$0.''14 \times 3.''47$ in order to accomplish the highest 
spectral resolution of $R\simeq 20000$. Thanks to the effective
adaptive optics system (where the target itself was used
as the guide star) enabling to reduce the size of stellar image 
down to FWHM~$\ltsim 0.''1$ irrespective of natural seeing condition, 
stellar photons could be efficiently collected even
in such a narrow slit. Actual exposures were done in two different
positions (A and B) by shifting the image in the direction of 
the slit length and one observation cycle consisted of an A-B-B-A 
sequence of four exposures. The time for one (A or B) 
exposure was from one second (at the shortest) to 15~minutes 
(at the longest) depending on the brightness. Several cycles were 
repeated for a star according to the necessity to achieve
a sufficient S/N ratio. For two very metal-poor stars of special
interest (BD+44$^{\circ}$493 and G~64-37), we expended comparatively 
long total exposure times of $\sim 1$~hr.

The reduction of the spectra (A$-$B subtraction for background
cancellation, flat-fielding, bad-pixel correction, cosmic-ray 
events correction, scattered-light subtraction, aperture 
extraction, wavelength calibration, co-adding of spectrum frames,
and continuum normalization) was performed by using the ``echelle'' 
package of the software IRAF\footnote{IRAF is distributed by the 
National Optical Astronomy Observatories, which is operated by 
the Association of Universities for Research in Astronomy, Inc. 
under cooperative agreement with the National Science Foundation.} 
in a standard manner.
For most of the targets, sufficiently high S/N ratios of 
$\sim$~100--200 (S/N~$\sim 300$ for BD+44$^{\circ}$493) were 
eventually accomplished. We confirmed by inspecting the 
line width that the expected spectral resolving power of 
$R\sim 20000$ is actually attained.

\section{Stellar Parameters}

\subsection{Atmospheric Parameters}

Regarding the atmospheric parameters ($T_{\rm eff}$, $\log g$, 
$v_{\rm t}$, and [Fe/H]) of the program stars necessary for 
constructing model atmospheres and determining abundances, 
various published studies were consulted. 
In case where two or more choices were possible, we 
preferentially selected spectroscopically determined ones.
The finally adopted parameter values (with the references) 
are presented in table 1. 
As seen from the $T_{\rm eff}$ vs. $\log g$ diagram shown in
figure 1a, our targets are roughly divided, according
to the surface gravity, into dwarfs ($\log g > 3$) and giants 
($\log g < 3$). Also noted from this figure is the trend
of $\log g$ tending to be lower with decreasing $T_{\rm eff}$.

\subsection{Kinematic Properties}

In order to examine the kinematic properties of the program stars, 
we computed their orbital motions within the galactic gravitational
potential based on the positional and proper-motion data 
(taken from SIMBAD database) along with the radial-velocity 
data (measured from our spectra), following the procedure 
described in subsection 2.2 of Takeda (2007).
The adopted input data and the resulting solutions of kinematic 
parameters (space velocity components, orbital eccentricity,
mean galactocentric radius, etc.) are summarized in electronic 
table E1. Figures 1b--d show the mutual correlations of
$z_{\rm max}$ (maximum separation from the galactic plane), 
$V_{\rm LSR}$ (tangential component of the space velocity 
relative to the Local Standard of Rest), and [Fe/H] (metallicity).
Applying Ibukiyama and Arimoto's (2002) classification criteria 
(cf. figure 1b), we can roughly divide our targets from 
these figures that $\sim 50\%$, $\sim 20\%$, and $\sim 30\%$ 
belong to halo population with [Fe/H]$\ltsim -1$,
thick-disk population with $-1 \ltsim$~[Fe/H]~$\ltsim -0.5$,
and thin-disk population with $-0.5 \ltsim$~[Fe/H],
respectively.

\setcounter{figure}{0}
\begin{figure}
  \begin{center}
    \FigureFile(90mm,90mm){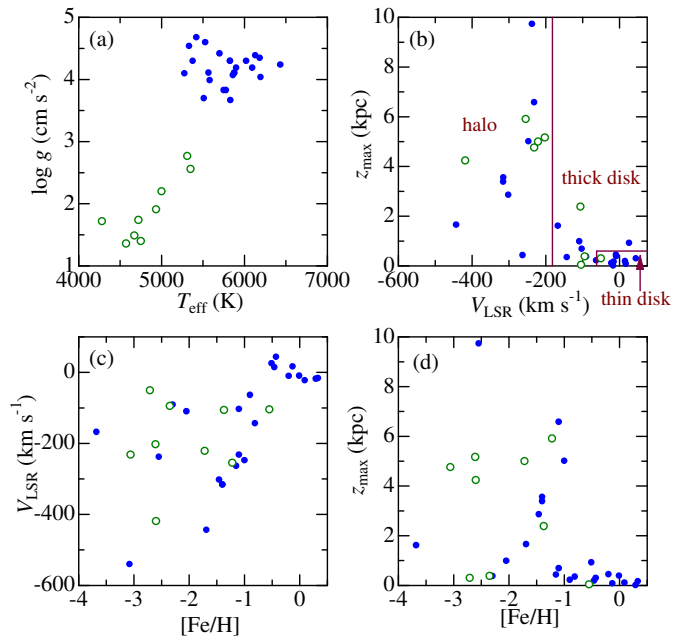}
  \end{center}
\caption{Correlations of representative stellar parameters. 
Filled (blue) and open (green) symbols indicate dwarfs ($\log g > 3$)
and giants ($\log g < 3$), respectively. 
(a) $T_{\rm eff}$ vs. $\log g$, (b) $V_{\rm LSR}$ vs. $z_{\rm max}$,
(c) [Fe/H] vs. $V_{\rm LSR}$, and (d) [Fe/H] vs. $z_{\rm max}$.
Note that, one star (G~64-37) is not plotted in panels (b) 
and (d), since its $z_{\rm max}$ turned out abnormally large 
and unreliable. 
}
\end{figure}

\section{Abundance Determination}

\subsection{Synthetic Spectrum Fitting}

We interpolated Kurucz's (1993) grid of ATLAS9 model 
atmospheres\footnote{These ATLAS9 models computed by Kurucz (1993)
approximately include the convective overshooting effect 
in an attempt to simulate the real convection as possible. 
It has been occasionally argued, however, that this treatment
may cause inconsistencies with observational quantities (e.g., 
colors or Balmer line profiles) and even the classical
pure mixing-length treatment ``without overshooting'' would be 
a better choice (e.g., Castelli et al. 1997). Since lines tend to 
become somewhat weaker in ``with overshooting'' atmospheres as 
compared to ``without overshooting'' cases because of the lessened 
temperature gradient in the lower part of the atmosphere,  some 
difference may be expected in resulting abundances between these 
two cases, especially for comparative higher $T_{\rm eff}$ stars
(i.e., early G to late A; cf. Fig. 24 of Castelli et al.) 
where the convection zone due to hydrogen ionization comes close 
to the bottom of the atmosphere.
We investigated how much S abundance difference would result in the 
analysis of S~{\sc i} 10455--10459 lines when ``without overshooting'' 
models were used instead of ``with overshooting'' ones. 
Test calculations for the representative case of 
$T_{\rm eff}$ = 6000~K and $\log g = 4.0$ (for [Fe/H] = 0 and $-3$)
revealed, however, that the abundance differences are only 
$\ltsim 0.05$~dex (i.e., slightly lower abundances are obtained
when the overshooting option is switched off). Accordingly, we may 
conclude that the difference of how the convection is treated 
is insignificant in the present case.} 
as well as the grid of the non-LTE departure 
coefficients computed by Takeda et al. (2005b) in terms of 
$T_{\rm eff}$, $\log g$, and [Fe/H] to generate the 
atmospheric model and the departure coefficient data 
for each star. 

Then we carried out non-LTE spectrum-synthesis analyses by 
applying Takeda's (1995) automatic fitting procedure to 
the region of S~{\sc i} 10455--10459 lines 
while regarding the sulfur abundance as well as 
the macro-broadening parameter and the radial velocity 
as adjustable parameters to be established.
The adopted atomic data of the relevant S~{\sc i} lines 
are presented in table 2.
How the theoretical spectrum for the converged solutions 
fits well with the observed spectrum is displayed in 
figure 2, and the resulting non-LTE S abundances ($A^{\rm N}$) 
are given in table 1.
Regarding the Sun, we obtained $A^{\rm N}_{\odot} = 7.20$
based on Kurucz et al.'s (1984) solar flux spectrum,
which was used as the reference solar sulfur abundance. 
Accordingly, we evaluated [S/H] and [S/Fe] values for each star
as [S/H] $\equiv A^{\rm N} - 7.20$ and 
[S/Fe] $\equiv$ [S/H] $-$ [Fe/H].

As to BD+44$^{\circ}$493, the most metal-poor star among our 
sample ([Fe/H]~$\sim -3.7$), we unfortunately could not detect 
any trace of S~{\sc i} lines in our spectrum (cf. figure 2) 
in spite of its fairly high S/N ratio ($\sim 300$).
We therefore estimated the upper limit of equivalent width 
for the strongest component at 10455.45~$\rm\AA$ as 
$EW_{10455}^{\rm UL} \simeq k \times$~FWHM/(S/N)~$\simeq 4$~m$\rm\AA$,
where $k$ is a factor we assumed to be 2 according to 
our experience and FWHM was set to 0.6~$\rm\AA$ (estimated 
from the width of the C~{\sc i} line at 10691~$\rm\AA$).
By using this $EW_{10455}^{\rm UL}$ and following the procedure
described in the next subsection, we derived
$A^{\rm N} \le 4.49$ (with $\Delta_{10455} = -0.34$) or 
[S/Fe]~$\le +0.96 (\simeq +1.0)$ for this star.

\subsection{Abundance-Related Quantities}

While the non-LTE synthetic spectrum fitting directly yields
the final abundance solution, this approach is not necessarily
suitable when one wants to evaluate the extent of non-LTE 
corrections or to study the abundance sensitivity to changing 
the atmospheric parameters (i.e., it is tedious to repeat 
the fitting process again and again for different assumptions 
or different atmospheric parameters).
Therefore, with the help of Kurucz's (1993) WIDTH9 program 
(which had been considerably modified in various respects; 
e.g., inclusion of non-LTE effects, etc.), we computed 
the equivalent widths for each of the triplet lines 
($EW_{10455}$, $EW_{10456}$, and $EW_{10459}$)  ``inversely'' 
from the abundance solution (resulting from non-LTE spectrum 
synthesis) along with the adopted atmospheric model/parameters, 
since they are much easier to handle.
Based on such evaluated $EW$ values, the LTE abundances for each of
the lines ($A^{\rm L}_{10455}$, $A^{\rm L}_{10456}$, and 
$A^{\rm L}_{10459}$) were freshly computed, from which the 
non-LTE corrections ($\Delta_{10455}$, $\Delta_{10456}$, and 
$\Delta_{10459}$) were derived such as 
$\Delta_{10455} \equiv A^{\rm L}_{10455} - A^{\rm N}$, etc.

We then estimated the uncertainties in $A^{\rm N}$
by repeating the analysis on $EW_{10455}$ ($EW$ for the strongest
component) while perturbing the standard values of atmospheric 
parameters interchangeably by $\pm 100$~K in $T_{\rm eff}$, 
$\pm 0.2$~dex in $\log g$, and $\pm 0.3$~km~s$^{-1}$ in 
$v_{\rm t}$ (which we regarded as typical uncertainties 
of the atmospheric parameters according to the original references). 
Let us call these six kinds of abundance variations as
$\delta_{T+}$, $\delta_{T-}$, $\delta_{g+}$, $\delta_{g-}$, 
$\delta_{v+}$, and $\delta_{v-}$, respectively.
We then computed the root-sum-square of three quantities
$\delta_{Tgv} \equiv (\delta_{T}^{2} + \delta_{g}^{2} + \delta_{v}^{2})^{1/2}$
as the abundance uncertainty (due to combined errors in 
$T_{\rm eff}$, $\log g$, and $v_{\rm t}$), 
where $\delta_{T}$, $\delta_{g}$, and $\delta_{v}$ are defined as
$\delta_{T} \equiv (|\delta_{T+}| + |\delta_{T-}|)/2$, 
$\delta_{g} \equiv (|\delta_{g+}| + |\delta_{g-}|)/2$, 
and $\delta_{v} \equiv (|\delta_{v+}| + |\delta_{v-}|)/2$,
respectively.\footnote{In addition to such estimated abundance 
ambiguities due to errors in the adopted atmospheric parameters, 
we should keep in mind that uncertainties caused by photometric 
random errors may become significant, especially for the case of
very weak lines near to the detection limit. For estimating these
errors, we may invoke the formula derived by Cayrel (1988),
who showed that the ambiguity in $EW$ is roughly expressed as 
$\sim 1.6 (w\;\delta x)^{1/2} \epsilon$, where $w$ is the typical 
line FWHM, $\delta x$ is the pixel size (in unit of wavelength), and
$\epsilon$ is the photometric accuracy represented by $\sim$~(S/N)$^{-1}$.
Substituting $w \sim$~0.5--1~$\rm\AA$, $\delta x \simeq 0.25 \rm\AA$, 
and $\epsilon \sim 1/100$, we obtain $\sim $~6--8~m$\rm\AA$ as 
the typical uncertainty in $EW$. This means that the S abundance 
derived for very metal-poor stars where the $EW$ of the strongest 
S~{\sc i} line ($EW_{10455}$) is around $\sim 10$~m$\rm\AA$  
could be subject to additional ambiguities due to photometric
errors which may amount to $\ltsim$~0.2--0.3~dex.} 

The resulting [S/Fe], $\Delta_{10455}$, and $EW_{10455}$ for 
each star are plotted against $T_{\rm eff}$ in figures 3a--c,
where the error bar attached to [S/Fe] represents $\delta_{Tgv}$.
We can see from these figures that 
$-0.4 \ltsim \Delta_{10455} \ltsim 0$
and 10~m$\rm\AA \ltsim EW_{10455} \ltsim 160$~m$\rm\AA$,
with a characteristic [Fe/H]-dependence differing between 
dwarfs and giants.
While only representative $A^{\rm N}$, $EW_{10455}$, $\Delta_{10455}$, 
and [S/Fe] are given in table 1, all the relevant data
(including $EW$ and $\Delta$ for all three lines and each of 
the $\delta$ values) are presented in electronic table E2.

\setcounter{figure}{2}
\begin{figure}
  \begin{center}
    \FigureFile(80mm,100mm){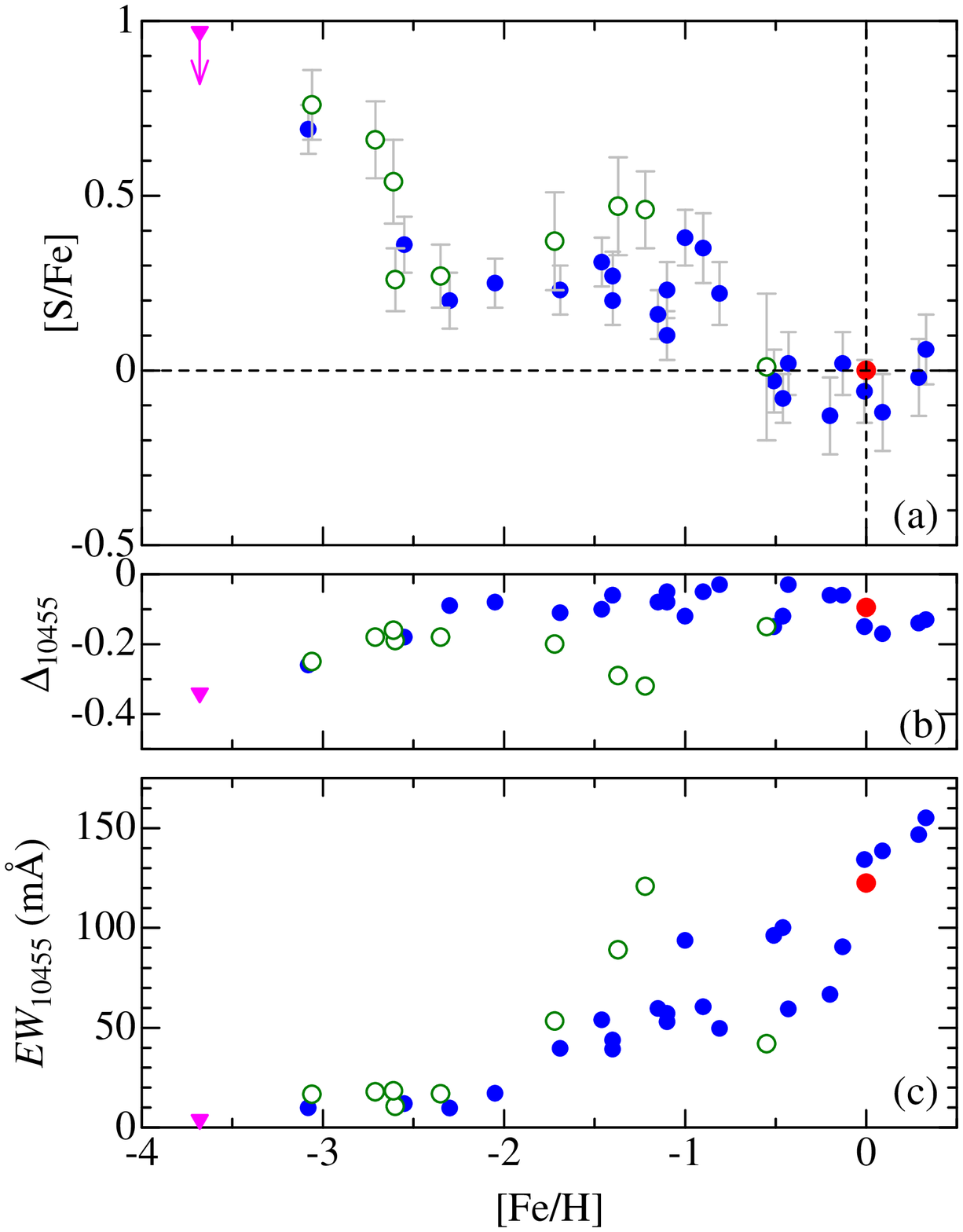}
  \end{center}
\caption{Sulfur abundances (along with the related quantities) 
plotted against [Fe/H]: 
(a) [S/Fe] (S-to-Fe logarithmic abundance ratio corresponding to non-LTE 
sulfur abundance; where attached error bars represent the ambiguities
due to uncertainties in the atmospheric parameters ($\delta_{Tgv}$;
cf. subsection 4.2)
(b) $\Delta_{10455}$ (non-LTE correction for the S~{\sc i}~10455 line). 
(c) $EW_{10455}$ (equivalent width for the S~{\sc i}~10455 line).
Dwarfs ($\log g > 3$) and giants ($\log g < 3$) are indicated by filled 
(blue) and open (green) symbols, respectively. The red filled circle 
denotes the Sun. For BD+44$^{\circ}$493, where 
the S~{\sc i} lines are invisible, the upper limit values corresponding 
to $EW_{10455} \le 4$~m$\rm\AA$ (cf. subsection 4.1) are shown by 
filled (pink) triangles.
}
\end{figure}

\section{Discussion}

\subsection{Behavior of [S/Fe] against [Fe/H]}

The [S/Fe] vs. [Fe/H] relation resulting from this investigation
is shown in figure 3a. A significant characteristic read from this 
figure is that the [S/Fe] ratio attains considerably high values 
($\sim$~+0.7--0.8~dex) in very metal-poor halo stars 
with [Fe/H]~$\sim -3$ (such as G~64-37, BD$-18^{\circ}$5550,
and HD~115444). The upper limit of [S/Fe] ($\le +1.0$) for 
BD+44$^{\circ}$493 ([Fe/H]~$\simeq -3.7$) does not contradict
this argument. It should be noted that this is a result 
after taking into account the negative (downward) non-LTE 
correction by $\sim$~0.2--0.3 dex.\footnote{Along with the non-LTE
correction, we should also pay attention to the 3D correction,
since our analysis is based on Kurucz's (1993) 1D model atmospheres.
Caffau et al. (2007) theoretically investigated the 3D effect 
on S abundance determinations for FGK population I dwarfs 
(including the Sun), and found that the 3D corrections for 
the solar S~{\sc i} 10455-10459 lines are on the order of 
+0.1~dex (cf. their table 2). Meanwhile, according to the recent
work of Caffau et al. (2010), the 3D corrections for the S 
abundances of metal-poor stars derived from S~{\sc i} 10455-10459 
lines are again $\sim +0.1$~dex (cf. their table 4). Therefore,
we can reasonably assume that the 3D effect is insignificant
for [S/Fe] (where ``relative'' abundance between a star and 
the Sun is involved) because of being canceled.} Accordingly, we conclude
that the behavior of [S/Fe] is different from the ``flat'' trend 
exhibited by other refractory $\alpha$-elements (e.g., Mg, Si, Ca, Ti) 
showing only moderately supersolar [$\alpha$/Fe] ratio of 
$\sim$~+0.3--0.5~dex at the metal-deficient regime between 
[Fe/H]~$\sim -2$ and $\sim -4$ (e.g., Cayrel et al. 2004). 
Such a difference in [$\alpha$/Fe] behavior at the very 
low-metallicity region between refractory and volatile 
species may pose a significant observational constraint on 
the chemical evolution in the early-time of the Galaxy (e.g., 
a necessity of including the effect of time-delayed deposition).

It should be stressed here, however, that [S/Fe] never simply 
keeps rising up to such a large value with a decrease 
in [Fe/H], as previously argued. The behavior of [S/Fe] at 
$-3 \ltsim$~[Fe/H]~$\ltsim 0$ we found here is actually 
more complicated:
After a gradual rise from [Fe/H]$\sim 0$ to $\sim -1$
(the typical tendency seen in $\alpha$ elements), 
a local plateau (or even a slight downward bending) is recognized
at $-2.5 \ltsim$~[Fe/H]~$\ltsim -1.5$ with a mildly supersolar 
[S/Fe] ($\sim +0.3$), which then experiences a sudden jump 
between the narrow interval at $-3 \ltsim$~[Fe/H]~$\ltsim -2.5$ 
up to [S/Fe]~$\sim$~+0.7--0.8 at [Fe/H]~$\sim -3$.
Thus, to say the least, the behavior of [S/Fe] is rather intricate
with a zigzag appearance, which is neither globally flat nor 
monotonically increasing. This is the consequence of this study. 

\subsection{Comparison with the Results from Different Lines}

In order to examine whether or not any systematic effect exists
between the abundances derived from different lines,
we compared the [S/H]$_{104}$ values derived 
from 10455--10459 lines in this study with
[S/H]$_{92}$ (9212/9228/9237 lines) or
[S/H]$_{86}$ (8693--4 lines) taken from the published 
results where data for any of our targets are available.
In doing this, we paid special attention to the consistency that 
both [S/H]'s to be compared correspond to the same atmospheric 
parameters ($T_{\rm eff}$, $\log g$, and $v_{\rm t}$). 
Regarding [S/H]$_{92}$ (non-LTE values), the data for 5 stars 
(G 64-37, G~29-23, G~18-39, HD~194598, HD~193901) were taken 
from table 1 of Nissen et al. (2007b), while those for 6 stars 
(HD~122563, HD~140283, HD~108317, HD~19445, HD~201891, HD~148816) 
are the non-LTE reanalysis results of Takada-Hidai et al.'s (2005) 
equivalent widths as done by Takeda et al. (2005b; cf. subsection 4.1
therein). Meanwhile, [S/H]$_{86}$ for 9 stars (HD~5682, HD~142373,
HD~165908, HD~10700, HD~131156, HD~141004, HD~196755, HD~161797,
HD~182572) were taken from Takeda (2007). Although these
[S/H]$_{86}$'s are LTE values, we used them as they are, 
since non-LTE corrections for 8693--4 lines are practically
negligible; i.e., only a few hundredths dex in the case of 
population I dwarfs (cf. Takeda et al. 2005b).
The resulting [S/H]$_{92}^{\rm NLTE}$ vs. [S/H]$_{104}^{\rm NLTE}$ 
and [S/H]$_{86}^{\rm LTE}$ vs. [S/H]$_{104}^{\rm NLTE}$ correlation 
plots are shown in figures 4a and b, respectively.  
We can see from these figures that the agreement is fairly good 
in most cases, though exceptionally large deviation is seen
at the low abundance end for each panel (G~64-37 and HD~6582).

\setcounter{figure}{3}
\begin{figure}
  \begin{center}
    \FigureFile(60mm,120mm){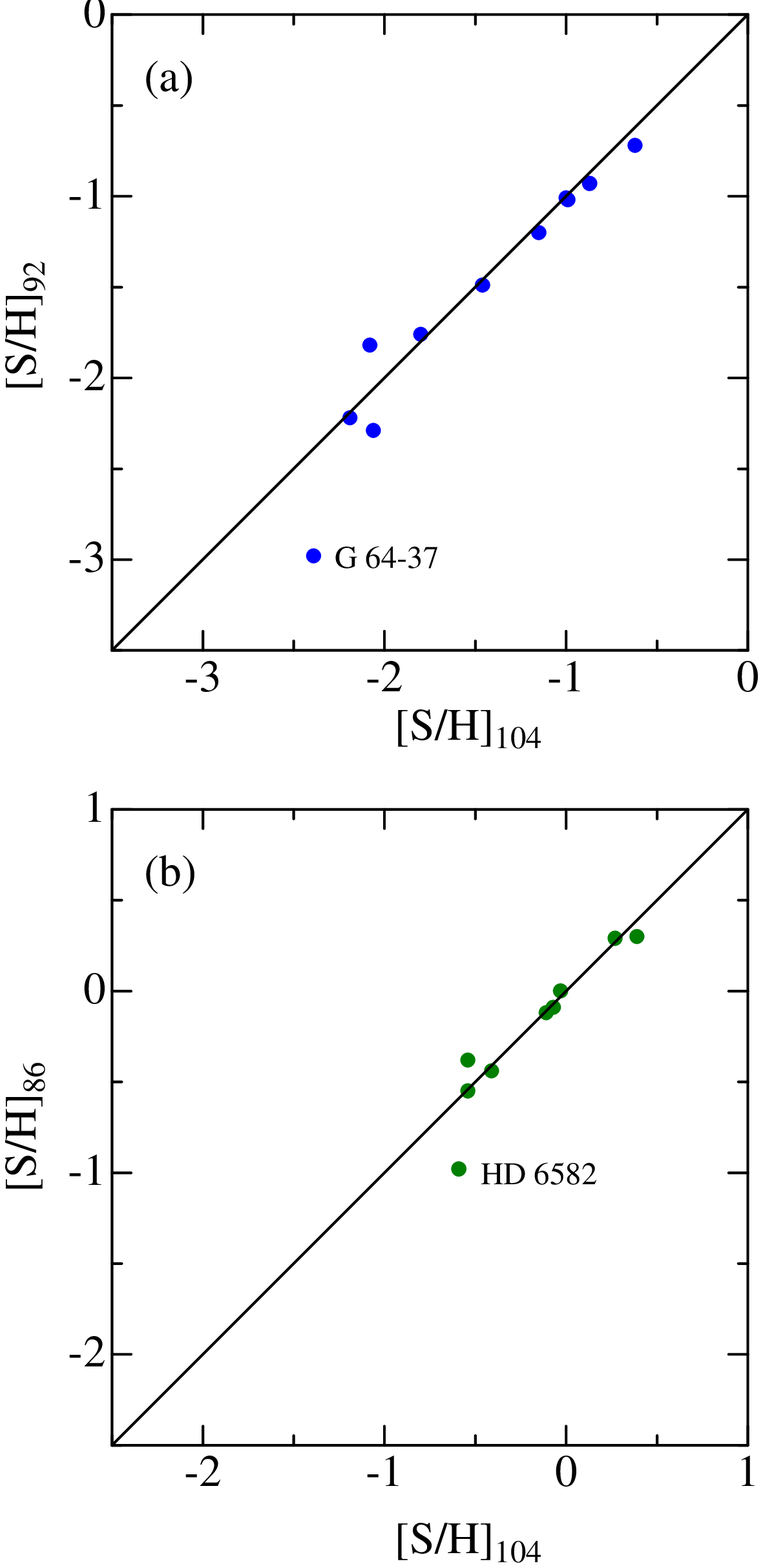}
  \end{center}
\caption{Correlation of ``star$-$Sun'' sulfur abundance difference determined 
in this study based on S~{\sc i}~10455--10459 lines of multiplet 3 
([S/H]$_{104}$) with those derived from other S~{\sc i} lines 
with the same atmospheric parameters as adopted in this study.
(a) Comparison with [S/H]$_{92}$ derived from S~{\sc i}~9212/9228/9237 lines 
of multiplet 2 for 11 stars in common (based on the data published by 
Takada-Hidai et al. 2005 and Nissen et al. 2007b)
(b) Comparison with [S/H]$_{86}$ derived from S~{\sc i}~8693--4 lines 
of multiplet 6 for 9 stars in common (based on the results obtained by 
Takeda 2007).
See subsection 5.2 for more details.}
\end{figure}

Regarding HD~6582, the reason for the disagreement 
($\sim 0.4$~dex) between [S/H]$_{104}^{\rm NLTE}$ 
($-0.59$; this study) and [S/H]$_{86}^{\rm LTE}$ ($-0.98$; Takeda 2007) 
is well understandable. This is due to the fact that the 
S~{\sc i} 8693--4 feature for this star is too weak (because of 
its low metallicity nature) to yield a reliable S abundance, 
as can be recognized from figure 4 of Takeda (2007). 
Therefore, [S/H]$_{86}$ should be regarded as erroneous for 
this case of HD~6582. Considering the satisfactory agreement
between [S/H]$_{86}^{\rm LTE}$ and [S/H]$_{104}^{\rm NLTE}$ for 
the other stars, we may state that S~{\sc i} 10455--10459 
lines can be a useful S abundance indicator also for comparatively
metal-rich population I stars, if non-LTE corrections are properly
taken into account.

On the other hand, the large discrepancy amounting to $\sim 0.6$~dex
shown by of G~64-37 ([S/H]$_{92}^{\rm NLTE} = -2.98$ and 
[S/H]$_{104}^{\rm NLTE} = -2.39$) is more serious and puzzling. 
Admittedly, our abundance result for this star may be subject to 
comparatively large uncertainty because its spectrum quality is 
not sufficiently good for measuring very weak lines (cf. figure 2). 
However, we consider that at least its detection is probably real; 
so the abundance of this star being lower by $\sim 0.6$~dex 
is rather hard to accept. At any rate, if Nissen et al.'s (2007b) 
low-scale result is correct, it has an important implication
that [S/Fe] as low as $\sim +0.1$ does exist at [Fe/H]~$\sim -3$, 
which does not match our consequence. 

Given the general agreement between [S/Fe]$_{104}^{\rm NLTE}$
and [S/Fe]$_{92}^{\rm NLTE}$,  we consider that 
the flat (or slightly bending) trend of [S/Fe]$_{92}^{\rm NLTE}$ 
at $\sim +0.2$ derived by Nissen et al. (2007b; cf. the lower panel 
of their figure 11) is reasonable (at least at 
$-2.7 \ltsim$~[Fe/H]~$\ltsim -1$), in the sense that it may correspond 
to the local plateau we found for [S/Fe]$_{104}^{\rm NLTE}$
at $-2.5 \ltsim$~[Fe/H]~$\ltsim -1.5$.
However, their low [S/Fe]$_{92}^{\rm NLTE}$ results of 
$\sim $~+0.1--0.3 for all three stars at [Fe/H]~$\sim -3$ are difficult 
to interpret, because of being discordant with what we found from 
[S/Fe]$_{104}^{\rm NLTE}$.  
Depending on cases, it might be necessary to abandon the idea 
that almost unique [S/Fe] should correspond to a given [Fe/H].
Namely, we may have to regard that [S/Fe] at this very 
low metallicity region is considerably diversified or bifurcated
in the sense that stars with high- and low-scale [S/Fe] are
mixed  around the same [Fe/H]. As a matter of fact, 
Caffau et al. (2005) suggested that [S/Fe] vs. [Fe/H] relation 
may be bimodal in the sense that it consists of two separate 
sequences branching off around [Fe/H]~$\sim -1.5$
(see also Caffau et al. 2010). If this is really the case,
it might suggest a difference in the degree of ISM mixing; 
i.e., changing from incompletely to well mixed with increasing 
[Fe/H] (e.g., Argast et al. 2000)

We do not insist, of course, that the behavior of [S/Fe] we concluded 
in this investigation (figure 3a) exclusively describes the nature of 
sulfur abundances of metal-poor regime applicable to all halo stars. 
It is clear that our object sample is still too small to extract
any decisive conclusion.
Obviously, further more S abundance studies on a much larger 
number of metal-deficient stars in a wide range of metallicity 
($-3 \ltsim$~[Fe/H]~$\ltsim -1$), preferably by using both 
S~{\sc i} 10455--10459 as well as 9212/9228/9237 lines, 
would be required to settle the problem.

\section{Conclusion}

Motivated by the recent debate on the [S/Fe] vs. [Fe/H] relation
of metal-deficient stars at $-3 \ltsim$~[Fe/H]~$\ltsim -1$,
where ``rising'' or ``flat'' tendencies have been differently 
suggested depending on the lines used (S~{\sc i} 8693--4 lines of 
multiplet 6 or S~{\sc i} 9212/9228/9237 lines of multiplet 1),
we conducted an extensive abundance analysis on 33 disk/halo stars 
in the wide metallicity range ($-3.6 \ltsim $~[Fe/H]~$\ltsim +0.3$)
by using S~{\sc i} 10455--10459 lines of multiplet 3,
a new probe potentially effective for exploring S abundances of 
very metal-poor stars.

The observations were carried out in 2009 July by using IRCS+AO188 
of the Subaru Telescope and spectra of sufficiently high resolution
($R \sim 20000$) and high S/N ($\sim$~100--200) were obtained 
in $zJ$-band (1.04--1.19~$\mu$m). Thanks to the high quality of
the data, we could successfully determine the sulfur abundances 
from the S~{\sc i} 10455--10459 triplet for most of the targets
by using the non-LTE spectrum synthesis technique.

We found an evidence of considerably large [S/Fe] ratio amounting 
to $\sim$~+0.7--0.8~dex at very low metallicity ([Fe/H]~$\sim -3$),
which makes a marked contrast with other refractory $\alpha$-elements 
(such as Mg, Si, Ca, Ti), which are known to show a flat tendency
at [$\alpha$/Fe]~$\sim 0.3$~over the whole halo metallicity range.
Given that S is a volatile element among the $\alpha$ group, such 
a difference may pose a significant constraint on the galactic 
chemical evolution.

The resulting global nature of [S/Fe] over the wide metallicity 
range is not so simple as has been argued (i.e., neither
simply flat nor ever increasing) but actually rather complex: 
After a gradual rise from [S/Fe] $\sim 0$ ([Fe/H]$\sim 0$) to 
[S/Fe] $\sim +0.3$ ([Fe/H]$\sim -1$), a local plateau (or even 
a slight downward bending) extends over 
$-2.5 \ltsim$~[Fe/H]~$\ltsim -1.5$ with a mildly supersolar 
[S/Fe] ($\sim +0.3$), which is followed by a sudden increase
between the narrow interval ($-3 \ltsim$~[Fe/H]~$\ltsim -2.5$) 
up to [S/Fe]~$\sim$~+0.7--0.8 at [Fe/H]~$\sim -3$.

We could not find any systematic difference between the abundances 
derived from these S~{\sc i} 10455--10459 lines and those from
other 8693--4 or 9212/9228/9237 lines, which are mostly 
in agreement. We thus consider that the flat trend of [S/Fe] 
at $\sim +0.2$ concluded by Nissen et al. (2007b) based on 9212/9237
lines actually correspond to the local plateau we found 
at $-2.5 \ltsim$~[Fe/H]~$\ltsim -1.5$. However, it is not clear 
why they did not detect such large [S/Fe] of $\sim$~0.7--0.8 
as we found at [Fe/H] $\sim -3$. One possible explanation
might be that stars with high- and low-scale [S/Fe] are
mixed around the same [Fe/H], though it has to be
observationally confirmed based on a much larger sample.

\bigskip

We express our heartful thanks to Y. Minowa and T.-S. Pyo
for their kind advices and helpful support in preparing
as well as during the IRCS+AO188 observations.

One of the authors (M. T.-H.) is grateful for a financial support 
from a grant-in-aid for scientific research (C, No. 22540255) 
from the Japan Society for the Promotion of Science.

This research has made use of the SIMBAD database, operated by
CDS, Strasbourg, France. 

\appendix
\section*{Double-Star Nature of HD~219617}

In our observation by using the Subaru Telescope with IRCS+AO188, 
we accidentally realized that HD~219617 was a double-star 
with similar components, thanks to the high spatial resolution 
($\ltsim 0.''1$) accomplished by the adaptive optics system (AO188). 
Consulting the SIMBAD database, 
we learned that this is really a double system (WDS~23171$-$1349) of 
$0.''8$ separation, comprising 9.08 ($V$) and 8.77 ($V$) F8~IV stars.
Since the fainter one is in the North-East direction and the brighter
one is in the South-East, we call the former and the latter
HD~219617 (NE) and HD~219617 (SW), respectively.
We obtained its $J$-, $H$-, and $K$-band images by making use of 
the imaging mode of IRCS (cf. figure 5), from which we confirmed
that NE is slightly fainter than SW also in near IR: 
$J_{\rm NE} - J_{\rm SW} = +0.20$,   
$H_{\rm NE} - H_{\rm SW} = +0.27$, and
$K_{\rm NE} - K_{\rm SW} = +0.20$.  

\setcounter{figure}{4}
\begin{figure}
  \begin{center}
    \FigureFile(80mm,120mm){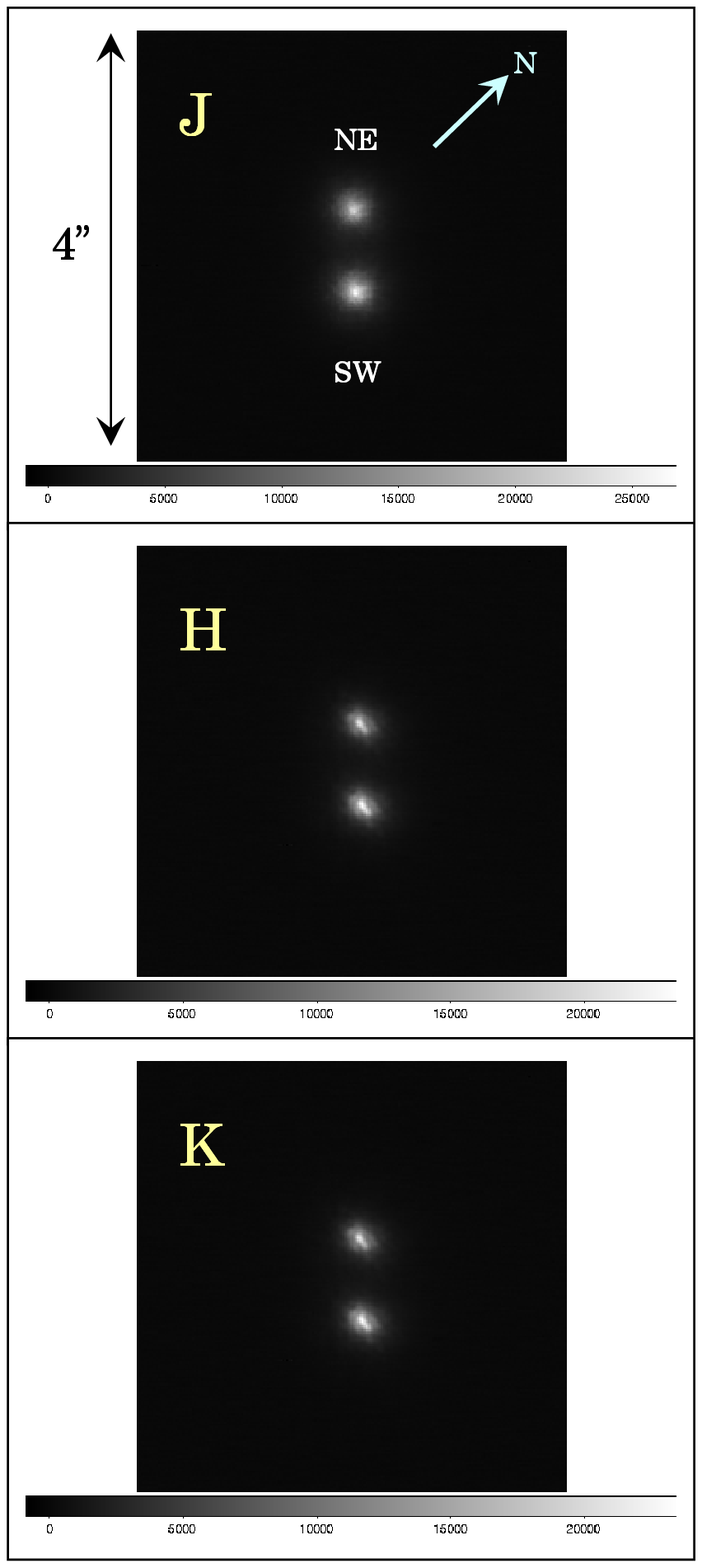}
  \end{center}
\caption{High-resolution images of HD~219617 in $J$, $H$, and $K$ bands,
which were obtained in the imaging mode of IRCS+AO188.}
\end{figure}

Although chemical abundances of this star have been repeatedly studied 
by a number of investigators, the combined spectrum of both components
seems to have been used without attending its binary nature, presumably
because these two tend to merge already in the natural seeing condition 
of $\sim 1''$--$2''$.
Since we could obtain the spectrum of each component separately,
we here briefly describe the comparison of these two.

A notable characteristic is that two spectra are remarkably 
similar to each other, as shown in figure 6. 
Actually, they are hardly discernible by eyes. 
According to the comparison of $EW$s (from $\sim 10$ to 
$\sim 230$~m$\rm\AA$) measured for 32 lines (of C~{\sc i}, Mg~{\sc i}, 
Si~{\sc i}, S~{\sc i}, Fe~{\sc i}, and Sr~{\sc ii}), we found that 
the correlation is very good ($r = 0.94$) and a linear regression 
analysis yielded
$EW_{\rm SW} = 1.020 EW_{\rm NE} - 1.3$ 
(where $EW$ is in unit of m$\rm\AA$). Thus, almost no essential 
systematic difference exists in terms of the line strengths.
Besides, the striking similarity of Paschen $\gamma$ wing (figure 6b) 
suggests that $T_{\rm eff}$ is almost the same. We thus assigned 
the same atmospheric parameters for both NE and SW. 
We also point out that the radial velocity is again practically the 
same ($V_{\rm rad}^{\rm hel}$ is +14~km~s$^{-1}$ and +13~km~s$^{-1}$ for
NE and SW, respectively; cf. electronic table E1).

\setcounter{figure}{5}
\begin{figure}
  \begin{center}
    \FigureFile(80mm,80mm){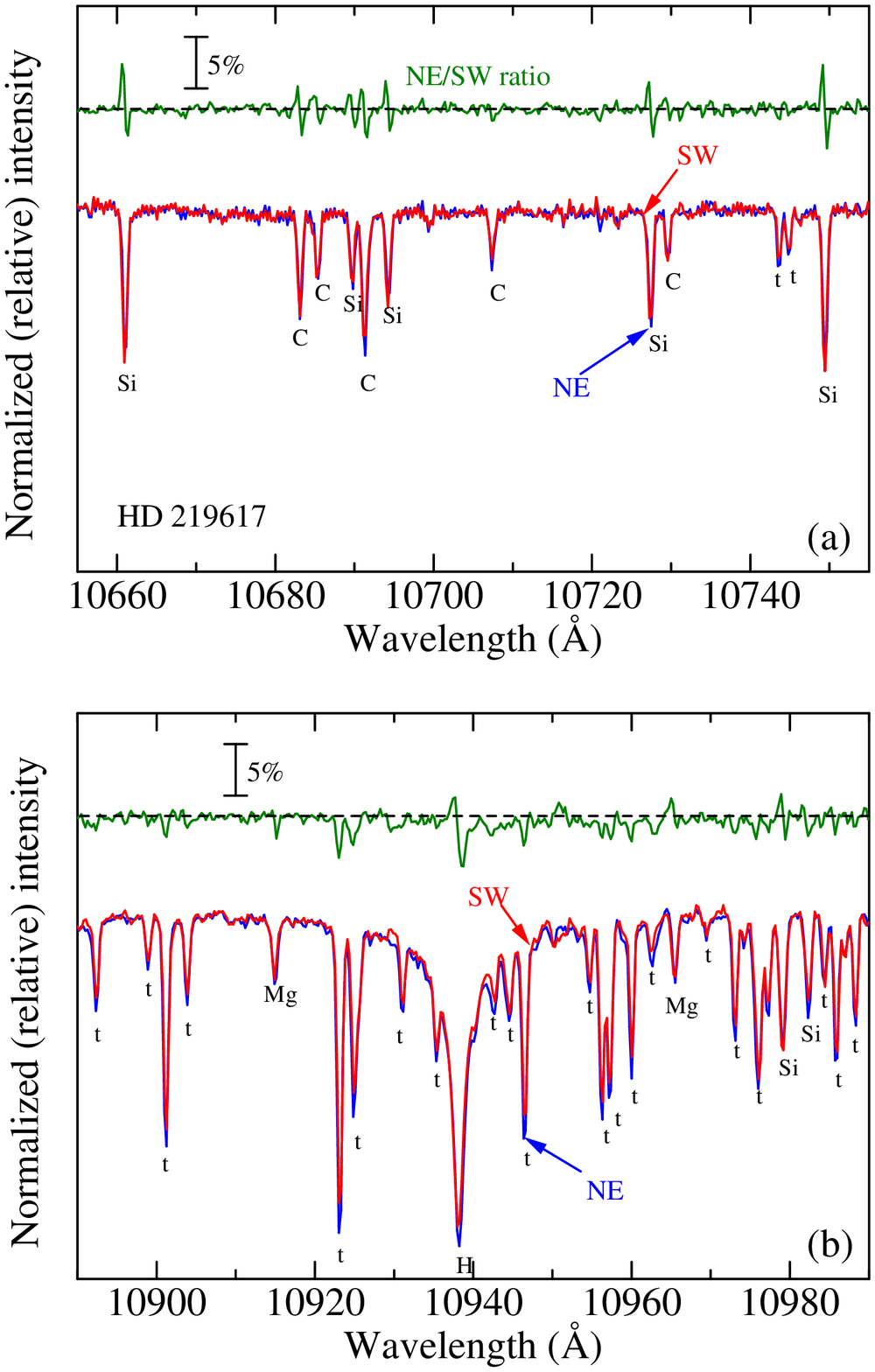}
  \end{center}
\caption{
Comparison of the spectra of HD~219617(NE) and HD~219617(SW).
(a) 10655--10755 $\rm\AA$ region comprising C~{\sc i} and Si~{\sc i} lines, 
(b) 10890--10990 $\rm\AA$ region comprising the conspicuous Paschen $\gamma$
line of neutral hydrogen. Though these two spectra are depicted in different
colors (NE in blue and SW in red), they are hardly discernible when overlapped.
The NE/SW spectrum residuals (NE divided by SW) are also displayed in each 
panel to demonstrate their remarkable similarity.
Note that many lines in panel (b) are telluric water vapor lines (denoted 
as ``t''). The wavelength scale of each spectrum has been adjusted to the
laboratory frame.
}
\end{figure}

Accordingly, previous studies having analyzed the combined spectrum
of HD~219617 as if it is a single star should have obtained almost
the correct result, because each of Sp(NE), Sp(SW), Sp(NE+SW) are 
essentially the same. Yet, one point to notice is, that stellar
parameters should not be derived from the luminosity simply 
estimated from the apparent (total) brightness of this double star.
Fulbright (2000), who evidently treated this star as being single,
initially guessed its $\log g$ from Hipparcos parallax (i.e., via
luminosity) as 3.9, while he finally adopted the spectroscopically
determined $\log g$ of 4.3. We speculate that this difference of
0.4~dex may have stemmed from the overestimation of the luminosity.

\clearpage
\twocolumn 
\setcounter{figure}{1}
\begin{figure}
  \begin{center}
    \FigureFile(150mm,200mm){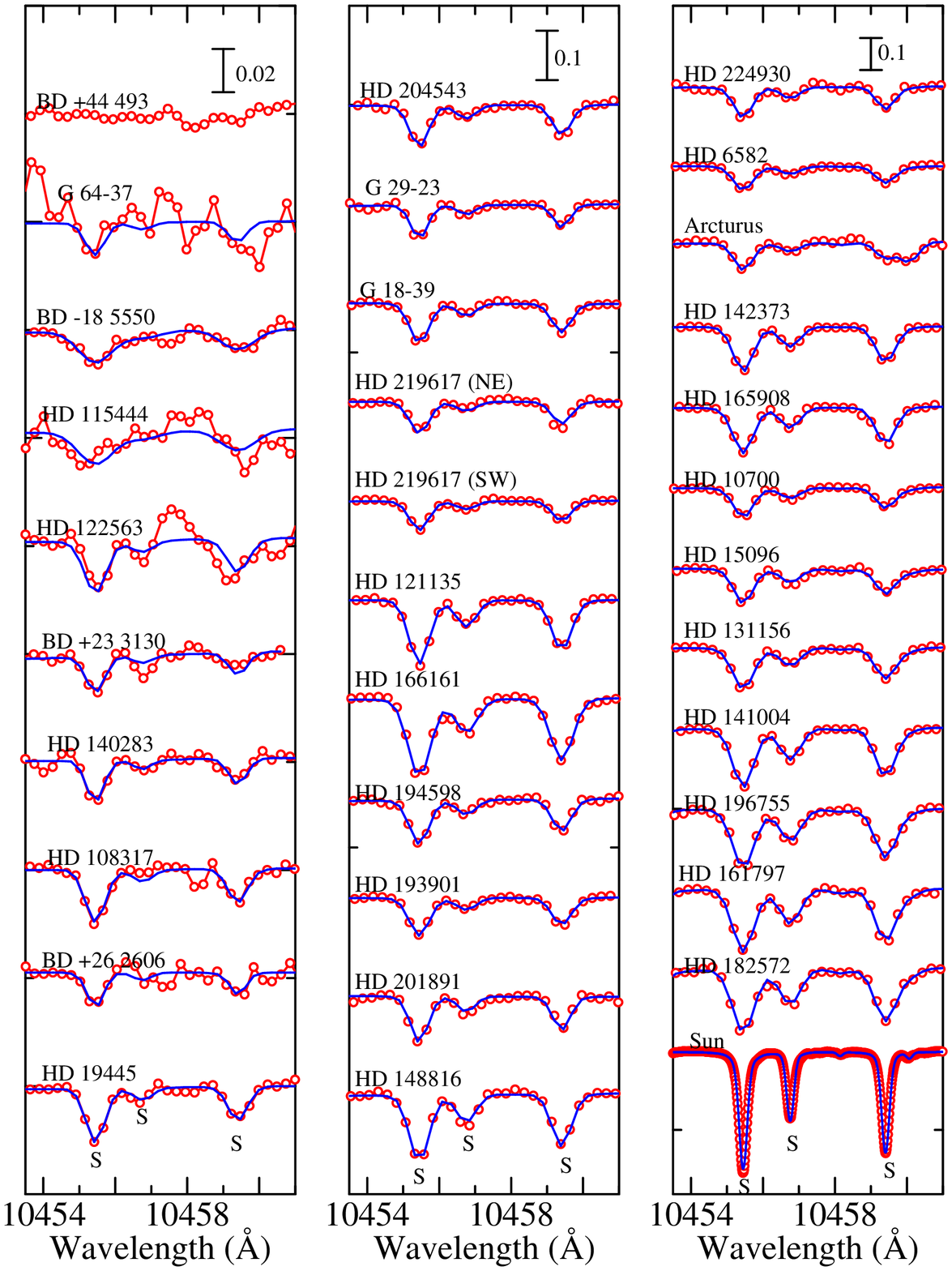}
  \end{center}
\caption{Synthetic spectrum fitting for the S~{\sc i} 10455--10459
triplet lines. The best-fit theoretical spectra are shown by (blue) 
solid lines, while the observed data are plotted by (red) symbols.  
In each panel (from left to right), the spectra are arranged 
(from top to bottom) in the ascending order of [Fe/H] 
as in table 1. An appropriate vertical offset (0.05, 0.2, 0.25 for 
the left, middle, and right panel, respectively) is applied 
to each spectrum relative to the adjacent one. 
The wavelength scale of each spectrum has been adjusted to the
laboratory frame. 
}
\end{figure}

\clearpage
\setcounter{table}{0}
\begin{table}[h]
\caption{Parameters of the program stars and the results of abundance analyses.}
\small
\begin{center}
\begin{tabular}{cc@{ }c@{ }c@{ }c c crccl} 
\hline\hline
Name & $T_{\rm eff}$ & $\log g$ & $v_{\rm t}$ & [Fe/H] & Ref. & $A^{\rm N}$ & 
$EW_{10455}$ & $\Delta_{10455}$ & [S/Fe] & Remark \\
 & (K) & (cm~s$^{-2}$) & (km~s$^{-1}$) & (dex) & & (dex) & (m$\rm\AA$) &(dex)&(dex)& \\
\hline
BD+44$^{\circ}$493 & 5510& 3.70 &1.30& $-$3.68 & ITO09 & 4.49& $(\le 4)$&($-$0.34)& $(\le +1.0)$ & S lines unmeasurable\\
G~64-37     & 6432& 4.24 &1.50& $-$3.08 & NIS07 & 4.81&   9.9& $-$0.26& +0.69& larger uncertainty\\
BD$-$18$^{\circ}$5550 & 4750& 1.40 &1.80& $-$3.06 & CAY04 & 4.90&  16.7& $-$0.25& +0.76& \\
HD~115444   & 4721& 1.74 &2.00& $-$2.71 & SIM04 & 5.15&  17.9& $-$0.18& +0.66& \\
HD~122563   & 4572& 1.36 &2.90& $-$2.61 & SIM04 & 5.13&  18.4& $-$0.16& +0.54& \\
BD+23$^{\circ}$3130 & 5000& 2.20 &1.40& $-$2.60 & FUL00 & 4.86&  10.6& $-$0.19& +0.26& \\
HD~140283   & 5830& 3.67 &1.90& $-$2.55 & MTH05 & 5.01&  12.0& $-$0.18& +0.36& \\
HD~108317   & 5310& 2.77 &1.90& $-$2.35 & MTH05 & 5.12&  16.9& $-$0.18& +0.27& \\
BD+26$^{\circ}$2606 & 5875& 4.10 &0.40& $-$2.30 & FUL00 & 5.10&   9.8& $-$0.09& +0.20& \\
HD~19445    & 6130& 4.39 &2.10& $-$2.05 & MTH05 & 5.40&  17.2& $-$0.08& +0.25& \\
HD~204543   & 4672& 1.49 &2.00& $-$1.72 & SIM04 & 5.85&  53.3& $-$0.20& +0.37& \\
G~29-23     & 6194& 4.04 &1.50& $-$1.69 & NIS07 & 5.74&  39.6& $-$0.11& +0.23& \\
G~18-39     & 6093& 4.19 &1.50& $-$1.46 & NIS07 & 6.05&  54.0& $-$0.10& +0.31& \\
HD~219617(NE) & 5825& 4.30 &1.40& $-$1.40 & FUL00 & 6.07&  43.8& $-$0.06& +0.27& North-East component\\
HD~219617(SW) & 5825& 4.30 &1.40& $-$1.40 & FUL00 & 6.00&  39.3& $-$0.06& +0.20& South-West component\\
HD~121135   & 4934& 1.91 &1.60& $-$1.37 & SIM04 & 6.30&  89.0& $-$0.29& +0.47& \\
HD~166161   & 5350& 2.56 &2.25& $-$1.22 & SIM04 & 6.44& 120.8& $-$0.32& +0.46& \\
HD~194598   & 6020& 4.30 &1.40& $-$1.15 & NIS07 & 6.21&  59.6& $-$0.08& +0.16& \\
HD~193901   & 5699& 4.42 &1.20& $-$1.10 & NIS07 & 6.33&  53.0& $-$0.05& +0.23& \\
HD~201891   & 5900& 4.19 &1.40& $-$1.10 & MTH05 & 6.20&  57.2& $-$0.08& +0.10& \\
HD~148816   & 5860& 4.07 &1.60& $-$1.00 & MTH05 & 6.58&  93.7& $-$0.12& +0.38& \\
HD~224930   & 5275& 4.10 &1.05& $-$0.90 & FUL00 & 6.65&  60.5& $-$0.05& +0.35& \\
HD~6582     & 5331& 4.54 &0.73& $-$0.81 & TAK05 & 6.61&  49.6& $-$0.03& +0.22& \\
Arcturus   & 4281& 1.72 &1.49& $-$0.55 & TAK09 & 6.66&  42.0& $-$0.15& +0.01& \\
HD~142373   & 5776& 3.83 &1.26& $-$0.51 & TAK05 & 6.66&  96.2& $-$0.15& $-$0.03& \\
HD~165908   & 6183& 4.35 &1.24& $-$0.46 & TAK05 & 6.66& 100.2& $-$0.12& $-$0.08& \\
HD~10700    & 5420& 4.68 &0.66& $-$0.43 & TAK05 & 6.79&  59.4& $-$0.03& +0.02& \\
HD~15096    & 5375& 4.30 &0.80& $-$0.20 & FUL00 & 6.87&  66.7& $-$0.06& $-$0.13& \\
HD~131156   & 5527& 4.60 &1.10& $-$0.13 & TAK05 & 7.09&  90.5& $-$0.06& +0.02& \\
HD~141004   & 5877& 4.11 &1.17& $-$0.01 & TAK05 & 7.13& 134.2& $-$0.15& $-$0.06& \\
HD~196755   & 5750& 3.83 &1.23& +0.09 & TAK05 & 7.17& 138.5& $-$0.17& $-$0.12& \\
HD~161797   & 5580& 3.99 &1.11& +0.29 & TAK05 & 7.47& 146.7& $-$0.14& $-$0.02& \\
HD~182572   & 5566& 4.11 &1.07& +0.33 & TAK05 & 7.59& 155.1& $-$0.13& +0.06& \\
Sun        & 5780& 4.44 &1.00&  0.00 & $\cdots$ & 7.20& 122.4& $-$0.10&  0.00& \\
\hline
\end{tabular}
\end{center}
In columns 1 through 6 are given the star designation, 
effective temperature, logarithmic surface gravity, 
microturbulent velocity dispersion, Fe abundance relative to
the Sun, and key for the reference of atmospheric parameters:
ITO09 $\cdots$ Ito et al. (2009),  NIS07 $\cdots$ Nissen et al. (2007b),
CAY04 $\cdots$ Cayrel et al. (2004), SIM04 $\cdots$ Simmerer et al. (2004),
FUL00 $\cdots$ Fulbright (2000), MTH05 $\cdots$ Takada-Hidai et al. (2005),
TAK05 $\cdots$ Takeda et al. (2005a), TAK09 $\cdots$ Takeda et al. (2009). 
Columns 7--10 present the results of the abundance analysis.
$A^{\rm N}$ is the non-LTE logarithmic abundance of S 
(in the usual normalization of H = 12.00) derived from spectrum-synthesis 
fitting, $EW_{10455}$ is the equivalent width (in m$\rm\AA$) for the 
S~{\sc i} 10455 line inversely computed from $A^{\rm N}$, $\Delta_{10455}$ 
is the non-LTE correction ($\equiv A^{\rm N} - A^{\rm L}_{10455}$) for the 
S~{\sc i} 10455 line, and [S/Fe] ($\equiv A^{\rm N}$ $-$ 7.20 $-$ [Fe/H]) 
is the S-to-Fe logarithmic abundance ratio relative to the Sun. 
Since HD~219617 is a double-star system ($0.''8$ separation) 
comprising two very similar stars, we assigned the same atmospheric 
parameters to both components (cf. the Appendix).
The objects are arranged in the ascending order of [Fe/H]. 
\end{table}

\clearpage
\setcounter{table}{1}
\begin{table}[h]
\caption{Atomic data of S~{\sc i} 10455--10458 triplet lines.}
\begin{center}
\begin{tabular}
{ccrrrrrr}\hline \hline
Mult. No. & Transition & $\lambda$ ($\rm\AA$) & $\chi$ (eV) & $\log gf$ & Gammar & Gammas &
Gammaw \\
\hline
3  & 4s $^{3}{\rm S}^{\rm o}_{1}$ -- 4p $^{3}{\rm P}_{2}$ & 10455.45 & 6.86 &  +0.26 & 8.86 & $-5.21$ & $-7.57^{*}$ \\
3  & 4s $^{3}{\rm S}^{\rm o}_{1}$ -- 4p $^{3}{\rm P}_{0}$ & 10456.76 & 6.86 &  $-0.43$ & 8.86 & $-5.21$ & $-7.57^{*}$ \\
3  & 4s $^{3}{\rm S}^{\rm o}_{1}$ -- 4p $^{3}{\rm P}_{1}$ & 10459.41 & 6.86 &  +0.04 & 8.86 & $-5.21$ & $-7.57^{*}$ \\
\hline
\end{tabular}
\end{center}
Note.\\
These data are were taken from Kurucz and Bell's (1995) compilation.
In the last three columns are given the damping parameters in the c.g.s. unit:
Gammar is the radiation damping constant, $\log\gamma_{\rm rad}$.
Gammas is the Stark damping width per electron density
at $10^{4}$ K, $\log(\gamma_{\rm e}/N_{\rm e})$.
Gammaw is the van der Waals damping width per hydrogen density
at $10^{4}$ K, $\log(\gamma_{\rm w}/N_{\rm H})$. \\
$^{*}$ Computed as default values in the Kurucz's WIDTH program
(cf. Leusin \& Tepil'skaya 1985).
\end{table}

\end{document}